\documentclass[a4paper,11pt]{article}
\usepackage{pos}

\def\lmatrix{\left(\begin{array}}
\def\rmatrix{\end{array}\right)}
\def\bea{\begin{eqnarray}}
\def\eea{\end{eqnarray}}
\def\nn{\nonumber}
\def\msbar{\overline{\rm MS\kern-0.5pt}\kern0.5pt}
\def\gsim{\mathrel{\rlap{\lower4pt\hbox{\hskip1pt$\sim$}}\raise1pt\hbox{$>$}}}
\def\lsim{\mathrel{\rlap{\lower4pt\hbox{\hskip1pt$\sim$}}\raise1pt\hbox{$<$}}}
\def\rho{\varrho}
\def\eps{\varepsilon}

\title{Mesonic decay constant and mass ratios and the conformal window}

\author[a]{Hee Sok Chung}
\author*[bc]{Daniel Nogradi}

\affiliation[a]{Korea University, Department of Physics, Seoul 02841, Korea}

\affiliation[b]{Eotvos University, Department of Theoretical Physics, Budapest 1117, Hungary}

\affiliation[c]{University of Adelaide, Adelaide, SA 5005, Australia}

\emailAdd{neville@korea.ac.kr}
\emailAdd{nogradi@bodri.elte.hu}

\abstract{Two particular ratios related to mesons are proposed for the study of the conformal window in
$SU(3)$ gauge theory and fundamental fermions. Lattice and other studies indicate that the lower end,
$N_f^*$, is at around 7 - 13 flavors which is a wide range without a clear consensus. Here we propose the decay
constant to mass ratios of mesons, $f_{PS,V} / m_V$, as a proxy since below the conformal window lattice
studies have shown that they are largely $N_f$-independent while at the upper end of the conformal window
they are vanishing. The drop from the non-zero constant value to zero at $N_f = 16.5$ might be indicative
of $N_f^*$. We compute $f_V / m_V$ to N$^3$LO and $f_{PS} / m_V$ to NNLO order in (p)NRQCD. The results
are unambiguously reliable just below $N_f = 16.5$, hence the results are expanded \'a la Banks-Zaks in
$\eps = 16.5 - N_f$. The convergence properties of the series and matching with the non-perturbative
infinite volume, continuum and chiral extrapolated lattice results at $N_f = 10$ suggest that the
perturbative results might be reliable down to $N_f = 12$. A sudden drop is observed at $N_f = 12$ and
$N_f = 13$ in $f_V / m_V$ and $f_{PS} / m_V$, respectively.}

\FullConference{The 40th International Symposium on Lattice Field Theory (Lattice 2023)\\
July 31st - August 4th, 2023\\
Fermi National Accelerator Laboratory\\}

\begin{document}
\maketitle

\section{Introduction and summary}

The lower end of the conformal window, usually denoted by $N_f^*$, for a given gauge group and fermion
representation, has been an elusive object of study
\cite{Appelquist:1988yc, Cohen:1988sq, Sannino:2004qp, Dietrich:2006cm,Appelquist:2007hu,Armoni:2009jn,Appelquist:2009ty,Fodor:2009wk, Frandsen:2010ej,Fodor:2011tu,Hasenfratz:2011xn,Aoki:2012eq,Fodor:2016zil,Hasenfratz:2016dou,Nogradi:2016qek,Fodor:2017gtj,Hasenfratz:2017qyr, Kim:2020yvr, Lee:2020ihn,Rummukainen:2022ekh}
. Naively, one would think the
lattice approach would be an ideal way to study it because once the infinite volume, continuum and chiral
limits are taken at each flavor $N_f$ below the loss of asymptotic freedom, the results would be 
unambiguous with a similarly unambiguous conclusion about $N_f^*$. However it became clear that
systematic effects close to the conformal window are significantly larger than at lower flavor numbers
where the models are very similar to QCD. As a result there is not a clear consensus for $SU(3)$ and
fundamental fermions, lattice results and non ab initio methods estimate $N_f^*$ to be somewhere in
the range $7 - 13$ which is rather broad.

In this contribution a new approach is proposed \cite{Chung:2023mgr}: 
by matching the fully controlled lattice results obtained for
low fermion numbers \cite{Nogradi:2019iek, Nogradi:2019auv, Kotov:2021mgp}
and fully controlled perturbative results obtained close to but below
$N_f = 16.5$ where asymptotic freedom is still present. The latter calculation will be the focus
of our contribution. The former are available in the range $2 \leq N_f \leq 10$ and the task is then to
study how far down the perturbative results can be trusted from $N_f = 16.5$ and if they can be matched
to the last non-perturbatively obtained point at $N_f = 10$. The particular quantities to be investigated
are dimensionless and finite ratios related to bound states; the decay constant to mass ratios of
mesons. More precisely the ratios $f_V / m_V$ and $f_{PS} / m_V$ will be
investigated in (p)NRQCD \cite{Caswell:1985ui,Bodwin:1994jh,Pineda:1997bj, Brambilla:1999xf, Brambilla:2004jw}
which is the appropriate framework inside the conformal window. These are
readily available in the range $2 \leq N_f \leq 10$ from past lattice studies either directly, as for
$f_{PS} / m_V$, or indirectly by using the KSRF relations \cite{Kawarabayashi:1966kd, Riazuddin:1966sw}
for $f_V / m_V$.

The setup of the perturbative calculation is as follows. We start from a CFT close to but below
$N_f = 16.5$ which is weakly coupled as shown by Banks-Zaks \cite{Banks:1981nn}. 
All particles are of course massless and
correlation functions fall off algebraically. A flavor singlet mass term is introduced which leads to
bound states whose masses and decay constants are proportional to $m^\alpha$ with the same exponent
$\alpha = 1 / ( 1 + \gamma)$ related to the mass anomalous dimension $\gamma$
\cite{DelDebbio:2010ze,DelDebbio:2010jy}. The constant of
proportionality can be computed perturbatively following the (p)NRQCD prescription. In the NRQCD
language all $N_f$ flavors are ``heavy'' and there are no ``light'' flavors and we need to keep the purely
perturbative terms only. The ratio of decay constants
and meson masses are then obtained as a series in the coupling with coefficients depending on $N_f$. In
the final step both the coupling and any explicit $N_f$ dependence is expanded in $\eps = 16.5 - N_f$
leading to constant coefficients. The final series obtained in this way contains both powers of $\eps$
and its logarithm.

As always with a perturbative result its reliability or convergence properties are non-trivial. For the
case of $f_V / m_V$ we have N$^3$LO results and a comparison of the NNLO and N$^3$LO results show that it
might be reliable down to $N_f = 12$. We assign a theoretical error by taking the difference between the
last two available orders. The lattice result for $f_V$ is not available directly, only for $f_{PS}$ but
we utilise the KSRF relation to estimate $f_V = \sqrt{2} f_{PS}$ on the range $2 \leq N_F \leq 10$.
Curiously, the perturbative result at $N_f = 12$ is compatible with the last non-perturbative result at
$N_f = 10$ within errors. Assuming a monotonous behavior the following picture emerges: 
$f_V / m_V$ is constant outside the conformal window and drops sharply at around $N_f = 12$ finally
reaching zero at $N_f = 16.5$. The sudden drop might be indicative of the lower end of the conformal
window.

A similar analysis for the other ratio, $f_{PS} / m_V$, could not be fully carried out because the
perturbative result is only available to NNLO order for $f_{PS}$. Nonetheless assuming the convergence
properties are similar to $f_V / m_V$ we are able to conclude similarly that the perturbative result might 
be reliable down to $N_f = 12$. A sudden drop in the ratio $f_{PS} / m_V$ seems to occur at around $N_f =
13$.

The perturbative calculation can be viewed in one of two ways. First, as alluded to above, it may be
thought of as perturbative (p)NRQCD without any terms containing $\Lambda_{QCD}$ explicitly. Or, it is
also instructive to view it as (p)NRQED with more diagrams due to the non-abelian nature of the
interaction. If viewed this way the bound states in question are analogous to the positronium. 
This view is useful because it is easy to see that all decay constants and meson masses will
be proportional to the fermion mass, just as the decay constants and masses of positronium are
proportional to the electron mass. There is no other scale in QED than the electron mass, and there is no
other scale than the fermion mass in our calculation either since we started from a CFT. The constant of
proportionality in QED is well-known to contain powers of $\alpha$ as well as $\log(\alpha)$ which in our
case will lead to $\eps$ and $\log(\eps)$. 

\section{Perturbative results}

As with any perturbative calculation a running scale $\mu$ is introduced and since we start from a CFT
the natural scale is $\mu = m$, the mass of the fermions\footnote{One could choose any scale $\mu > m$}. 
All results will be given in the $\msbar$
scheme and the renormalized coupling will be denoted by 
$g^2(\mu)/(16\pi^2) = g^2(m)/(16\pi^2) = a$. The decay constants
and meson masses are expanded in $a$ leading to 
\cite{Penin:2004ay,Kniehl:2006qw,Bodwin:2008vp,Penin:2014zaa,Chung:2020zqc}
\bea
\label{fm}
f_{PS,V} &=& b_0 \, m \, a^{3/2} \left( 1 + b_{10} a + b_{11} a \log a + b_{20} a^2 + b_{21} a^2 \log a +
b_{22} a^2 \log^2 a + O(a^3) \right) \nn \\
m_V &=& c_0\, m \left( 1 + c_{20} a^2 + c_{30} a^3 + c_{31} a^3 \log a + O(a^4) \right) \;.
\eea
The leading term in the mass, $c_0 = 2$, just follows from having a free fermion and
anti-fermion pair whereas the first correction $c_{20} < 0$ is familiar from the quantum mechanical binding energy in a
Coulomb potential. Further radiative corrections are systematically obtained using (p)NRQCD. The leading
expression, $b_0$, for the decay constants is proportional to the ground state wave function at the origin. The
explicit form of the corrections, NNLO for $f_{PS}$ and N$^3$LO for $f_V, m_V$ can be found in
\cite{Chung:2023mgr}.

In the ratio $m$ drops out and 
the massless limit takes the running coupling to the fixed point $a(m) \to a_*$ which can be
expanded in $\eps = 16.5 - N_f$ and is known to 5-loops
\cite{Tarasov:1980au, Larin:1993tp, vanRitbergen:1997va, Czakon:2004bu, Baikov:2016tgj, Herzog:2017ohr},
\bea
\label{astar}
a_* = \eps\, ( e_0 + e_1\, \eps + e_2 \, \eps^2 + e_3 \, \eps^3 + \ldots )\,,
\eea
with some coefficients $e_i$.
Combining this expansion with (\ref{fm}) leads to the final results,
\bea
\label{fc}
\frac{f_V}{m_V} &=& \eps^{3/2} C_0 
\left( 1 + \sum_{n=1}^3 \sum_{k=0}^{n} C_{nk}\, \eps^n \log^k\eps + O(\eps^4) \right) \\
C_0    &=& 0.005826678 \nn \\
C_{10} &=& 0.4487893  \qquad C_{11} = - 0.2056075 \nn \\
C_{20} &=& 0.2444502  \qquad C_{21} = -0.1624891 \quad C_{22} = 0.03522870 \nn \\
C_{30} &=& 0.10604(3) \;\;\,\quad C_{31} = -0.1128420 \quad C_{32} = 0.03695458  \quad C_{33} = -0.005633665 \nn
\eea
for the vector case and
\bea
\label{fd}
\frac{f_{PS}}{m_V} &=& \eps^{3/2} C_0 
\left( 1 + \sum_{n=1}^2 \sum_{k=0}^{n} D_{nk}\, \eps^n \log^k\eps + O(\eps^3) \right) \\
D_{10} &=& 0.4654041  \quad D_{11} = -0.2056075 \nn \\
D_{20} &=& 0.2845697  \quad D_{21} = -0.1737620 \quad D_{22} = 0.03528692 \nn 
\eea
for the pseudo-scalar case.

Note that the coefficients in (\ref{fc}) and (\ref{fd}) are well-behaved as the order grows, in contrast to the 
generally factorially growing perturbative coefficients such as (\ref{fm}) and (\ref{astar}).
Furthermore, since $f_{PS,V} / m_V$ are finite, RG-invariant, scheme independent 
physical quantities, all coefficients in
(\ref{fc}) and (\ref{fd}) are scheme independent as well\footnote{The coefficient $C_{30}$ is only
available in numerical form with some uncertainty at the moment.}. These coefficients are the main 
results of our work.

\begin{figure}
\begin{center}
\includegraphics[width=7.5cm]{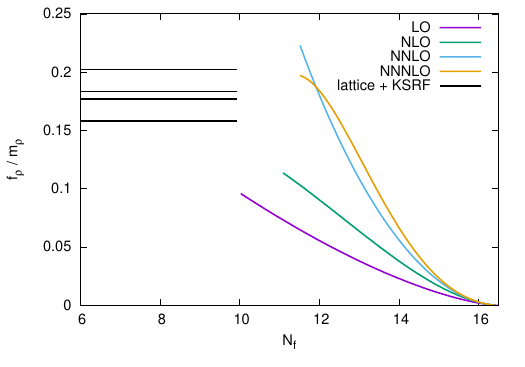} \includegraphics[width=7.5cm]{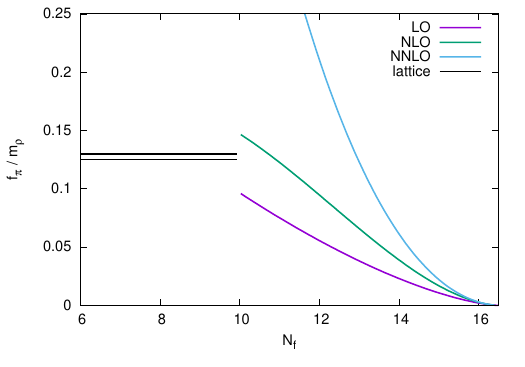}
\caption{Left: the $f_V / m_V$ ratio in increasing perturbative order. 
The non-perturbative result from combined lattice calculations
\cite{Nogradi:2019iek, Nogradi:2019auv, Kotov:2021mgp}
and the KSRF-relation is also shown. The smaller error band corresponds to the uncertainty of the lattice
calculation, the wider one combines this with a conservative estimate of the uncertainty of the
KSRF-relation itself. Right:
The corresponding results for $f_{PS} / m_V$. 
}
\label{pertplot}
\end{center}
\end{figure}

\section{Matching low $N_f$ and high $N_f$}

The increasing orders for the two ratios are shown in figure \ref{pertplot}. 
Clearly, the deviation between the NNLO and N$^3$LO results of $f_V / m_V$ for
$N_f \geq 12$ is not substantial. Quantitatively, in the range $11.9 \leq N_f \leq 12.1$, the
deviation between the NNLO and N$^3$LO results is at most 4\%, or in the range $11.5 \leq N_f \leq 12.5$
at most 13\%. We thus conclude that in the region of interest, $N_f \sim 12$, the N$^3$LO result is
robust and reliable. We take as an estimate of the neglected higher orders the difference between the
last two available orders. 

\begin{figure}
\begin{center}
\includegraphics[width=7.5cm]{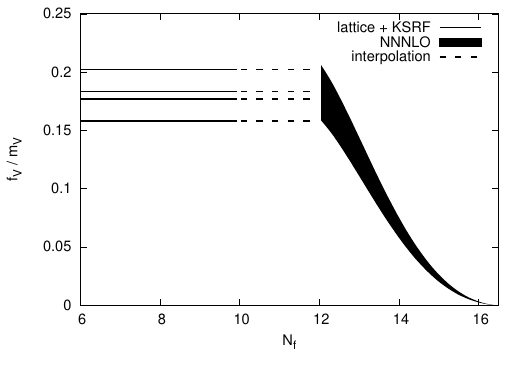} \includegraphics[width=7.5cm]{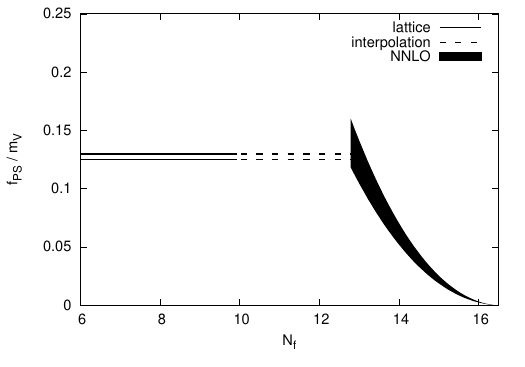} 
\end{center}
\caption{
    Non-perturbative lattice results in the range $2\leq N_f \leq 10$ and the perturbative ones 
    where they seem reliable. Assuming a monotonous behavior only a small range needs to be interpolated.
    Left: $f_V / m_V$, the wider error bands for $2\leq N_f \leq 10$ includes the error from the usage of
    the KSRF relation. The error of the perturbative curve is estimated from the
    difference of the last two available orders. 
    Right: $f_{PS} / m_V$, the error of the perturbative result is estimated from 
    that of $f_V / m_V$.
}
\label{interp}
\end{figure}

In order to compare with the lattice results we would need $f_V$, however only $f_{PS}$ was measured
directly \cite{Nogradi:2019iek, Nogradi:2019auv, Kotov:2021mgp}. 
Here we use the KSRF relation $f_V = \sqrt{2} f_{PS}$ originating in vector
meson universality to estimate $f_V$ and assign a $12\%$ uncertainty which holds in QCD.
Somewhat unexpectedly the perturbative result at $N_f = 12$ matches the 
last non-perturbative lattice result almost exactly; see left panel of figure \ref{interp}. 

A similar analysis unfortunately cannot be completed for $f_{PS} / m_V$ because $f_{PS}$ is only
available to NNLO order. One may nonetheless assume that the theoretical uncertainty is similar to that
of $f_V$ making a comparison with the lattice results at low $N_f$ feasible; see right panel of
\ref{interp}.

In both cases a match between the low $N_f$ and high $N_f$ regions seems plausible. Assuming a monotonous
behavior and trusting the lattice results below $N_f = 10$ and the perturbative ones above $N_f =12$ or
$13$ leaves only a narrow range to be interpolated or more rigorously calculated in future lattice work.
The following picture seems to emerge: a mostly $N_f$-independent flat curve drops sharply at around $N_f
= 12$ and $N_f = 13$ for the two ratios, respectively. This sudden change in behavior might be indicative
of $N_f^*$, the lower end of the conformal window.

\section{Conclusion and outlook}

In this contribution we presented a new approach to shed light on the emergence of conformal behavior from
chirally broken dynamics as the flavor number increases which combines both perturbative and
non-perturbative information. Well chosen dimensionless quantities were
presented which can be easily measured in lattice calculations in the chiral limit and which can also be
computed in perturbation theory, again in the massless limit. Current lattice calculations are able to
provide unambiguous results for low $N_f$ and the perturbative results are reliable at high $N_f \leq 16.5$.
Curiously, the two approaches seem to match at around $N_f = 12, 13$ where a sudden change in behavior as
a function of $N_f$ is observed.

The results can be improved in a number of ways. First, direct lattice results at $N_f = 12$ would be
very useful. The difficulty is controlling all 3 sources of systematic effects, finite volume, finite
lattice spacing and finite mass which certainly would lead to very costly calculations. Second, the
dominant source of uncertainty of $f_V / m_V$ was the use of the KSRF relation, which in principle could
be eliminated once $f_V$ is measured directly on the lattice. Third, currently the highest (p)NRQCD order
for $f_{PS}$ is NNLO which in principle could be extended to N$^3$LO, similarly to $f_V$. 
However going beyond N$^3$LO order for any quantity does not seem feasible in the near future since the 6-loop
$\beta$-function would be needed for that.


\begin{thebibliography}{99}

\footnotesize

\bibitem{Appelquist:1988yc}
T.~Appelquist, K.~D.~Lane and U.~Mahanta,
Phys. Rev. Lett. \textbf{61}, 1553 (1988)


\bibitem{Cohen:1988sq}
A.~G.~Cohen and H.~Georgi,
Nucl. Phys. B \textbf{314}, 7-24 (1989)


\bibitem{Sannino:2004qp}
F.~Sannino and K.~Tuominen,
Phys. Rev. D \textbf{71}, 051901 (2005)
[arXiv:hep-ph/0405209 [hep-ph]].


\bibitem{Dietrich:2006cm}
D.~D.~Dietrich and F.~Sannino,
Phys. Rev. D \textbf{75}, 085018 (2007)
[arXiv:hep-ph/0611341 [hep-ph]].


\bibitem{Appelquist:2007hu}
T.~Appelquist et al.,
Phys. Rev. Lett. \textbf{100}, 171607 (2008),
\textbf{102}, 149902 (2009)
[arXiv:0712.0609 [hep-ph]].


\bibitem{Armoni:2009jn}
A.~Armoni,
Nucl. Phys. B \textbf{826}, 328-336 (2010)
[arXiv:0907.4091 [hep-ph]].


\bibitem{Appelquist:2009ty}
T.~Appelquist, G.~T.~Fleming and E.~T.~Neil,
Phys. Rev. D \textbf{79}, 076010 (2009)
[arXiv:0901.3766 [hep-ph]].


\bibitem{Fodor:2009wk}
Z.~Fodor et al.,
Phys. Lett. B \textbf{681}, 353-361 (2009)
[arXiv:0907.4562 [hep-lat]].


\bibitem{Frandsen:2010ej}
M.~T.~Frandsen, T.~Pickup and M.~Teper,
Phys. Lett. B \textbf{695}, 231-237 (2011)
[arXiv:1007.1614 [hep-ph]].


\bibitem{Fodor:2011tu}
Z.~Fodor et al.,
Phys. Lett. B \textbf{703}, 348-358 (2011)
[arXiv:1104.3124 [hep-lat]].


\bibitem{Hasenfratz:2011xn}
A.~Hasenfratz,
Phys. Rev. Lett. \textbf{108}, 061601 (2012)
[arXiv:1106.5293 [hep-lat]].


\bibitem{Aoki:2012eq}
Y.~Aoki et al.,
Phys. Rev. D \textbf{86}, 054506 (2012)
[arXiv:1207.3060 [hep-lat]].


\bibitem{Fodor:2016zil}
Z.~Fodor et al.,
Phys. Rev. D \textbf{94}, no.9, 091501 (2016)
[arXiv:1607.06121 [hep-lat]].


\bibitem{Hasenfratz:2016dou}
A.~Hasenfratz and D.~Schaich,
JHEP \textbf{02}, 132 (2018)
[arXiv:1610.10004 [hep-lat]].


\bibitem{Nogradi:2016qek}
D.~Nogradi and A.~Patella,
Int. J. Mod. Phys. A \textbf{31}, no.22, 1643003 (2016)
[arXiv:1607.07638 [hep-lat]].


\bibitem{Fodor:2017gtj}
Z.~Fodor et al.,
Phys. Lett. B \textbf{779}, 230-236 (2018)
[arXiv:1710.09262 [hep-lat]].


\bibitem{Hasenfratz:2017qyr}
A.~Hasenfratz, C.~Rebbi and O.~Witzel,
Phys. Lett. B \textbf{798}, 134937 (2019)
[arXiv:1710.11578 [hep-lat]].


\bibitem{Kim:2020yvr}
B.~S.~Kim, D.~K.~Hong and J.~W.~Lee,
Phys. Rev. D \textbf{101}, no.5, 056008 (2020)
[arXiv:2001.02690 [hep-ph]].


\bibitem{Lee:2020ihn}
J.~W.~Lee,
Phys. Rev. D \textbf{103}, no.7, 076006 (2021)
[arXiv:2008.12223 [hep-ph]].


\bibitem{Rummukainen:2022ekh}
K.~Rummukainen and K.~Tuominen,
Universe \textbf{8}, no.3, 188 (2022)


\bibitem{Chung:2023mgr}
H.~S.~Chung and D.~Nogradi,
Phys. Rev. D \textbf{107}, no.7, 074039 (2023)
[arXiv:2302.06411 [hep-ph]].


\bibitem{Nogradi:2019iek}
D.~Nogradi and L.~Szikszai,
JHEP \textbf{05}, 197 (2019)
[erratum: JHEP \textbf{06}, 031 (2022)]
[arXiv:1905.01909 [hep-lat]].


\bibitem{Nogradi:2019auv}
D.~Nogradi and L.~Szikszai,
PoS \textbf{LATTICE2019}, 237 (2019)
[arXiv:1912.04114 [hep-lat]].


\bibitem{Kotov:2021mgp}
A.~Y.~Kotov et al.,
JHEP \textbf{07}, 202 (2021)
[erratum: JHEP \textbf{06}, 032 (2022)]
[arXiv:2107.05996 [hep-lat]].


\bibitem{Caswell:1985ui}
W.~E.~Caswell and G.~P.~Lepage,
Phys. Lett. B \textbf{167}, 437-442 (1986)


\bibitem{Bodwin:1994jh}
G.~T.~Bodwin et al.,
Phys. Rev. D \textbf{51}, 1125-1171 (1995),
 \textbf{55}, 5853 (1997)
[arXiv:hep-ph/9407339 [hep-ph]].


\bibitem{Pineda:1997bj}
A.~Pineda and J.~Soto,
Nucl. Phys. B Proc. Suppl. \textbf{64}, 428-432 (1998)
[arXiv:hep-ph/9707481 [hep-ph]].


\bibitem{Brambilla:1999xf}
N.~Brambilla, A.~Pineda, J.~Soto and A.~Vairo,
Nucl. Phys. B \textbf{566}, 275 (2000)
[arXiv:hep-ph/9907240 [hep-ph]].


\bibitem{Brambilla:2004jw}
N.~Brambilla, A.~Pineda, J.~Soto and A.~Vairo,
Rev. Mod. Phys. \textbf{77}, 1423 (2005)
[arXiv:hep-ph/0410047 [hep-ph]].


\bibitem{Kawarabayashi:1966kd}
K.~Kawarabayashi and M.~Suzuki,
Phys. Rev. Lett. \textbf{16}, 255 (1966)


\bibitem{Riazuddin:1966sw}
Riazuddin and Fayyazuddin,
Phys. Rev. \textbf{147}, 1071-1073 (1966)


\bibitem{Banks:1981nn}
T.~Banks and A.~Zaks,
Nucl. Phys. B \textbf{196}, 189-204 (1982)


\bibitem{DelDebbio:2010ze}
L.~Del Debbio and R.~Zwicky,
Phys. Rev. D \textbf{82}, 014502 (2010)
[arXiv:1005.2371 [hep-ph]].


\bibitem{DelDebbio:2010jy}
L.~Del Debbio and R.~Zwicky,
Phys. Lett. B \textbf{700}, 217-220 (2011)
[arXiv:1009.2894 [hep-ph]].


\bibitem{Penin:2004ay}
A.~A.~Penin et al.,
Nucl. Phys. B \textbf{699}, 183-206 (2004), \textbf{829}, 398-399 (2010)
[arXiv:hep-ph/0406175 [hep-ph]].


\bibitem{Kniehl:2006qw}
B.~A.~Kniehl et al.,
Phys. Lett. B \textbf{638}, 209-213 (2006)
[arXiv:hep-ph/0604072 [hep-ph]].


\bibitem{Bodwin:2008vp}
G.~T.~Bodwin, H.~S.~Chung, J.~Lee and C.~Yu,
Phys. Rev. D \textbf{79}, 014007 (2009)
[arXiv:0807.2634 [hep-ph]].


\bibitem{Penin:2014zaa}
A.~A.~Penin and N.~Zerf,
JHEP \textbf{04}, 120 (2014)
[arXiv:1401.7035 [hep-ph]].



\bibitem{Chung:2020zqc}
H.~S.~Chung,
JHEP \textbf{12}, 065 (2020)
[arXiv:2007.01737 [hep-ph]].


\bibitem{Tarasov:1980au}
O.~V.~Tarasov, A.~A.~Vladimirov and A.~Y.~Zharkov,
Phys. Lett. B \textbf{93}, 429-432 (1980)


\bibitem{Larin:1993tp}
S.~A.~Larin and J.~A.~M.~Vermaseren,
Phys. Lett. B \textbf{303}, 334-336 (1993)
[arXiv:hep-ph/9302208 [hep-ph]].


\bibitem{vanRitbergen:1997va}
T.~van Ritbergen et al.,
Phys. Lett. B \textbf{400}, 379-384 (1997)
[arXiv:hep-ph/9701390 [hep-ph]].


\bibitem{Czakon:2004bu}
M.~Czakon,
Nucl. Phys. B \textbf{710}, 485-498 (2005)
[arXiv:hep-ph/0411261 [hep-ph]].


\bibitem{Baikov:2016tgj}
P.~A.~Baikov et al.,
Phys. Rev. Lett. \textbf{118}, no.8, 082002 (2017)
[arXiv:1606.08659 [hep-ph]].


\bibitem{Herzog:2017ohr}
F.~Herzog, B.~Ruijl, T.~Ueda, J.~A.~M.~Vermaseren and A.~Vogt,
JHEP \textbf{02}, 090 (2017)
[arXiv:1701.01404 [hep-ph]].

\end{thebibliography}
\end{document}